# Are Neutrinos Their Own Antiparticles?[*]

**Boris Kayser**

Fermilab, MS 106, P.O. Box 500, Batavia, IL 60510, U.S.A.

Email: boris@fnal.gov

**Abstract:** We explain the relationship between Majorana neutrinos, which are their own antiparticles, and Majorana neutrino masses. We point out that Majorana masses would make the neutrinos very distinctive particles, and explain why many theorists strongly suspect that neutrinos do have Majorana masses. The promising approach to confirming this suspicion is to seek neutrinoless double beta decay. We introduce a toy model that illustrates why this decay requires nonzero neutrino masses, even when there are both right-handed and left-handed weak currents.

For given helicity $h$, is each neutrino mass eigenstate $\nu_i$ identical to its antiparticle, or different from it? Equivalently, do neutrinos have Majorana masses? If they do, then, as we shall explain, each $\nu_i$ is identical to its antiparticle: $\overline{\nu_i(h)} = \nu_i(h)$. Neutrinos of this nature are referred to as Majorana neutrinos, while ones for which $\overline{\nu_i(h)} \neq \nu_i(h)$ are called Dirac neutrinos.

Let us recall what a Majorana mass is. Out of, say, a left-handed neutrino field, $\nu_L$, and its charge conjugate, $\nu_L{}^c$, one can build the "left-handed" (so called because it is constructed from $\nu_L$) Majorana mass term

$$\mathcal{L}_L = m_L \overline{\nu_L} \nu_L{}^c, \qquad (1)$$

which absorbs a $(\overline{\nu})_R$ and creates a $\nu_L$. As this illustrates, Majorana neutrino masses mix neutrinos and antineutrinos, so they do not conserve the lepton number $L$ that is defined by $L(\nu) = L(\ell^-) = -L(\overline{\nu}) = -L(\ell^+)$, where $\ell$ is a charged lepton. Moreover, a Majorana mass for any fermion $f$ mixes $f$ with its antiparticle $\bar{f}$, or with a related antifermion having the same electric charge as $\bar{f}$. Thus, quark and charged-lepton Majorana masses are forbidden by electric charge conservation. However, neutrinos, being electrically neutral, are permitted to possess Majorana masses, and such masses would make them *very* distinctive. For, a Majorana neutrino mass cannot

---

arise from the neutrino analogue of the Standard-Model (SM) coupling that gives quarks and charged leptons their masses. That analogue would be a Yukawa coupling of the form $H_{SM}\overline{\nu_R}\nu_L$, where $H_{SM}$ is the SM Higgs field. Instead, Majorana masses must come from couplings such as $H_{SM}H_{SM}\overline{\nu_L^c}\nu_L$ or $H_{I_W=1}\overline{\nu_L^c}\nu_L$. The first of these leads to non-renormalizability, and so is outside the spirit of the SM. The second involves a Higgs boson with weak isospin $I_W=1$, and there is no such boson in the SM. In addition, one can have the right-handed Majorana mass $m_R\overline{\nu_R^c}\nu_R$, which doesn't involve any Higgs field at all. One way or another, Majorana neutrino masses must have a different origin than the masses of quarks and charged leptons.

To see why the neutrino mass eigenstates will be Majorana neutrinos if neutrinos have Majorana masses, we note first that the objects $\nu_L$ and $\nu_L^c$ that appear in the Majorana mass of Eqn. (1) are not the mass eigenstates, but just the neutrinos in terms of which our neutrino model is constructed. As we have noted, the Majorana mass of Eqn. (1) induces mixing between $\nu_L^c$ (a right-handed antineutrino) and $\nu_L$. Now, we recall that, as a result of $\overline{K^0} \leftrightarrow K^0$ mixing, the neutral $K$ mass eigenstates are not $K^0$ and $\overline{K^0}$, but $K_S$ and $K_L$. Neglecting CP violation, the latter particles are just the states

$$K_{S,L} = \left(K^0 \pm \overline{K^0}\right)/\sqrt{2}. \qquad (2)$$

Clearly, each of these states is self-conjugate (apart from an irrelevant sign) under particle-antiparticle interchange. In a similar way, as a result of the $\nu_L^c \leftrightarrow \nu_L$ mixing induced by the Majorana mass of Eqn. (1), the neutrino mass eigenstate that results from this mass term is

$$\nu_i = \nu_L + \nu_L^c. \qquad (3)$$

Clearly, this $\nu_i$ satisfies $\nu_i^c = \nu_i$. Thus, like $K_S$ or $K_L$, $\nu_i$ is identical to its antiparticle.

Among theorists, there is a widespread belief that neutrinos do have Majorana masses. One reason for this prejudice is the following argument: The electroweak SM may be defined by a few principles that include $SU(2)_L \times U(1)_Y$ gauge invariance and renormalizability, and by its field or particle content. The original version of the SM did not include neutrino masses, and, leaving these masses aside, it is observed that anything allowed by the defining SM principles actually occurs in nature. Now, if we extend the SM to include neutrino masses, we note that right-handed Majorana masses are allowed by the defining SM principles. Therefore, it seems likely that Majorana neutrino masses occur in nature too.

To determine whether Majorana masses do occur in nature, the most promising approach is to seek neutrinoless double beta decay ($0\nu\beta\beta$). This is the process $Nucl \rightarrow Nucl' + e^-e^-$, in which one nucleus decays into another plus two electrons. The observation of this decay at any nonzero level would imply the existence in nature of a neutrino Majorana mass term.[1] To see this, we note that at the quark level, $0\nu\beta\beta$ is the process $dd \rightarrow uu + e^-e^-$. If this process is observed, then, by crossing, the amplitude for the process $e^+\bar{u}d \rightarrow e^-u\bar{d}$ must be nonzero. The SM tells us that the amplitudes for all the (virtual) processes $(\bar{\nu})_R \to e^+ W^-$, $W^- \to \bar{u}d$, $u\bar{d} \to W^+$, and $e^- W^+ \to \nu_L$ are nonzero as well. Thus, combining amplitudes, we conclude that the amplitude for the chain $(\bar{\nu})_R \to e^+ W^- \to e^+(\bar{u}d) \to e^-(u\bar{d}) \to e^- W^+ \to \nu_L$ must be non-vanishing. But this chain results in $(\bar{\nu})_R \to \nu_L$, which is precisely the effect of the Majorana mass term of Eqn. (1). Hence, the observation of $0\nu\beta\beta$ would imply the existence of a non-vanishing amplitude that is equivalent to a Majorana mass term. Consequently, this observation would also imply that neutrinos are Majorana particles.

Although $0\nu\beta\beta$ can receive contributions from a variety of sources, one anticipates that it will be dominated by the diagram in Fig. 1. There, the vertices labeled "SM vertex" are SM charged-current vertices, $U$ is the leptonic mixing matrix, and, as indicated, the amplitude is a coherent sum over the contributions of all the neutrino mass eigenstates $\nu_i$.

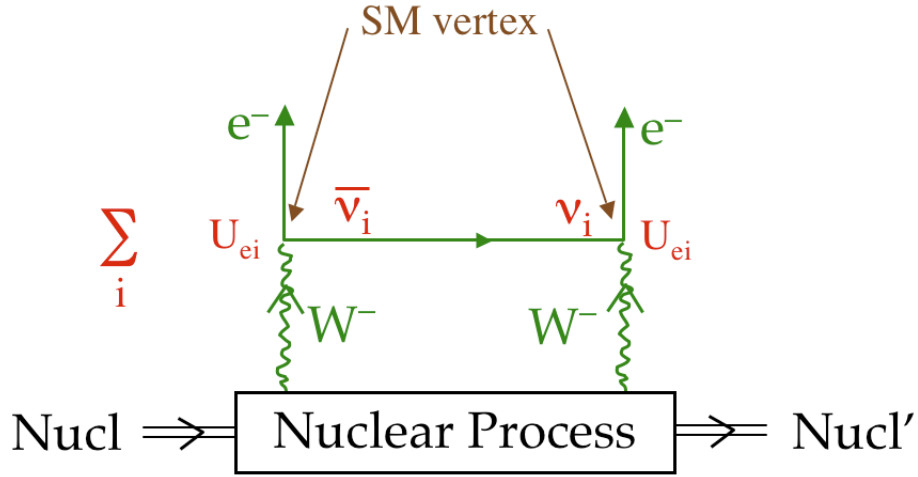

Figure 1. The diagram expected to dominate neutrinoless double beta decay.

The SM charged-current interaction that acts at each of the two leptonic vertices in Fig. 1 is lepton-number conserving. If the neutral particle that is absorbed at the vertex on the right is to create an electron, it must be a neutrino, not an antineutrino. However, at the vertex on the left, where this same particle is emitted, it is created by a W boson together with an electron, so at this vertex it must be an antineutrino, not a neutrino. Thus, the diagram in Fig. 1 does not exist unless $\bar{\nu}_i = \nu_i$. This is another way to see that the observation of $0\nu\beta\beta$ would imply that neutrinos are identical to their antiparticles.

Owing to the left-handed chiral character of the SM charged-current interaction, the "antineutrino" created by a $W$ boson together with an $e^-$ will be in a state that is dominantly of right-handed helicity, even if there is no difference between neutrinos and antineutrinos. Thus, the "$\bar{\nu}_i$" emitted at the leptonic vertex on the left in Fig. 1 is in a state that is mostly of right-handed helicity. However, if $m_i$ is the mass of $\nu_i$ and $E$ is its energy, this state does have a small component, of order $m_i/E$, with left-handed helicity. At the vertex on the right in Fig. 1, where the $\nu_i$ is absorbed to make an $e^-$, the SM left-handed charged current can absorb without suppression only its left-handed-helicity component. Thus, the contribution of $\nu_i$ exchange to the diagram of Fig. 1 is proportional to $m_i$.

Summing over $i$ and including the factors of $U_{ei}$ that appear at the vertices, we see that if the diagram of Fig. 1 dominates, then the amplitude for $0\nu\beta\beta$, Amp[$0\nu\beta\beta$], is proportional to the quantity

$$\left|\sum_i m_i U_{ei}^2\right| \equiv m_{\beta\beta}. \tag{4}$$

This quantity is known as the effective Majorana neutrino mass for neutrinoless double beta decay.

We see that when the diagram of Fig. 1 dominates, Amp[$0\nu\beta\beta$] is proportional to neutrino mass. Our discussion of helicities can leave one with the misimpression that this proportionality to mass is due merely to a mismatch of helicities at the two leptonic vertices. One might be tempted to believe that if the current acting at the leptonic vertex on the right in Fig. 1 were a non-SM right-handed (RH) current, rather than the SM left-handed (LH) one, then the diagram could lead to $0\nu\beta\beta$ without the need for any neutrino mass. Indeed, some years ago, it was common for an experiment obtaining an upper bound on the rate for $0\nu\beta\beta$ to quote its result as a bound on $m_{\beta\beta}$, and, alternatively, as a bound on the strength of any RH current, as if a RH current could engender $0\nu\beta\beta$ all by itself, without there being any neutrino mass. However, we now understand that, even if the current acting at one of the leptonic vertices in Fig. 1 is a RH current, $0\nu\beta\beta$ still requires nonzero neutrino mass. We have already seen that, without any assumption about the underlying mechanism driving $0\nu\beta\beta$, the observation of this decay would imply a Majorana neutrino mass. To see the need for mass in a way that is more specific to the mechanism in Fig. 1, we note first that $0\nu\beta\beta$ does not conserve lepton number $L$. Its initial state, a nucleus, has $L=0$. Its final state, containing another nucleus, two leptons, and no antileptons, has $L=2$. Now, as already mentioned, the SM interactions in Fig. 1 *do* conserve $L$. Moreover, a SM-like leptonic interaction that merely has a RH current in place of the SM LH current would conserve $L$ as well. Absent any non-SM $L$-violating interactions, the $\Delta L = 2$ of $0\nu\beta\beta$ can only come from Majorana neutrino masses such as the one of Eqn. (1). By absorbing a $\bar{\nu}$ and creating a $\nu$, this Majorana mass term causes the required $\Delta L = 2$. If we were to turn off the neutrino masses, including their Majorana masses, no lepton number violation would remain, and $0\nu\beta\beta$ would not occur.

The essential role of neutrino mass, even when there are RH currents, is nicely illustrated by a parity-conserving toy model that contains both LH and RH currents. In this model, we assume that there is only one generation. We suppose that the $W$ couples to the electron and neutrino fields via the parity-conserving interaction

$$-\mathcal{L}_W = \frac{g}{\sqrt{2}} W_\lambda^- \bar{e}\gamma^\lambda \nu + \text{h.c.} = \frac{g}{\sqrt{2}} W_\lambda^- \left(\overline{e_L}\gamma^\lambda \nu_L + \overline{e_R}\gamma^\lambda \nu_R\right) + \text{h.c.}, \tag{5}$$

where $g$ is a coupling constant. Finally, we suppose that the neutrino mass term is the $L \Leftrightarrow R$ symmetric

$$-\mathcal{L}_M = \frac{1}{2} m_M \left(\overline{\nu_L^c}\nu_L + \overline{\nu_R^c}\nu_R\right) + m_D \overline{\nu_R}\nu_L + \text{h.c.}. \tag{6}$$

Here, the Majorana mass $m_M$, and the "Dirac mass" $m_D$ (the neutrino analogue of the masses of other Dirac fermions such as the quarks), are both taken to be real and positive, and it is assumed that $m_M > m_D$.

In this model, there are two Majorana neutrino mass eigenstates:

$$v_2 = \frac{1}{\sqrt{2}}\left[(v_L + v_R) + (v_L + v_R)^c\right], \text{ with mass } m_2 = m_M + m_D, \qquad (7)$$

and

$$v_1 = \frac{1}{\sqrt{2}}\left[(v_L - v_R) + (v_L - v_R)^c\right], \text{ with mass } m_1 = m_M - m_D. \qquad (8)$$

In terms of them,

$$-\mathcal{L}_w = \frac{g}{\sqrt{2}} W_\lambda \left[\overline{e_L}\gamma^\lambda \frac{1}{\sqrt{2}}(v_{2L} + v_{1L}) + \overline{e_R}\gamma^\lambda \frac{1}{\sqrt{2}}(v_{2R} - v_{1R})\right] + \text{h.c.} \qquad (9)$$

To see what this toy model predicts for $0\nu\beta\beta$, let us consider the particle-physics part of this process, pictured in Fig. 2. The current that acts at the vertices of the diagram in Fig. 2 can be the LH current in Eqn. (5) at both vertices, or the RH current in Eqn. (5) at both vertices, or the LH current at one vertex and the RH current at the other. We find that if the LH current acts at both vertices, the amplitude for the process involves the neutrino masses through the factor

$$A_{LL} = \frac{m_2}{q^2 - m_2^2} + \frac{m_1}{q^2 - m_1^2} \cong \frac{m_2 + m_1}{q^2} = \frac{2m_M}{q^2}. \qquad (10)$$

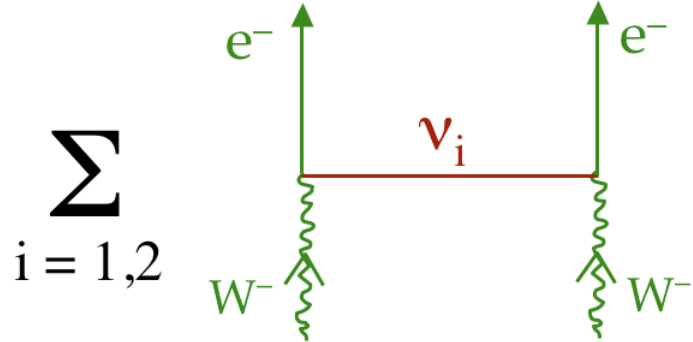

Figure 2. The particle-physics part of $0\nu\beta\beta$.

Here, q is the momentum transfer carried by the exchanged neutrino. In the second step in Eqn. (10) we have used the fact that in $0\nu\beta\beta$, $q \approx 50$ MeV $\gg m_{1,2}$, and in the last step we have used Eqns. (7) and (8).

We see that when the LH current acts at both vertices, the amplitude for $0\nu\beta\beta$ is proportional to the Majorana neutrino mass $m_M$. Fig. 3, in which neutrino mass is treated perturbatively, makes clear the reason for this. A Majorana mass turns a $\bar{\nu}$ into a $\nu$ (and a $\nu$ into a $\bar{\nu}$), and, like any mass term, reverses the handedness of the particle. Thus, an insertion of the Majorana mass $m_M$ is just what is needed to convert the *RH antineutrino* emitted together with an electron by the lepton-number-conserving LH current that acts at the vertex on the left in Fig. 3, into the *LH neutrino* that can be absorbed to make a second electron by the lepton-number-conserving LH current that acts at the vertex on the right.

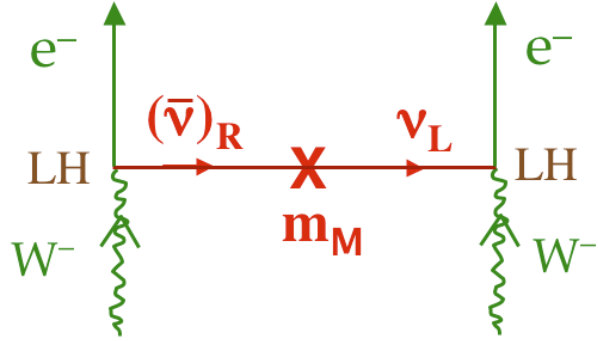

Figure 3. The particle-physics part of $0\nu\beta\beta$, with neutrino mass treated perturbatively, and LH currents acting at both vertices, as indicated.

When we go to higher order in neutrino mass, we can have multiple mass insertions in the neutrino line of Fig. 3. However, together these insertions must reverse the lepton number of the emitted antineutrino. Thus, there must be an odd number of insertions of the Majorana mass $m_M$. Together, the mass insertions must also reverse the handedness of the exchanged particle. Since both Majorana and Dirac masses reverse handedness, there must be an odd total number of mass insertions, including both $m_M$ and $m_D$. The implication of this argument is that if the exact expression for $A_{LL}$, Eqn. (10), is expanded in powers of $m_2$ and $m_1$, and the expansion is rewritten in terms of $m_M$ and $m_D$, every term must be of the form $m_M^{Odd\ Power} m_D^{Even\ Power}$. It is trivial to verify that this is indeed the case.

If the RH current in Eqn. (5) acts at both vertices in Fig. 2, the amplitude once again involves the neutrino masses through the same factor $A_{LL}$ as before, so that it is again proportional, to lowest order in the mass, to $m_M$.

Now suppose the LH current in Eqn. (5) acts at the vertex on the left in Fig. 2, but the RH current acts at the one on the right. This is the combination that was once thought capable of producing $0\nu\beta\beta$ without the need for any neutrino mass. However, we find that in our illustrative toy model, this combination of currents leads to an amplitude for the process in Fig. 2 that involves the neutrino masses through the factor

$$A_{LR} = i\rlap{/}{q}\left(\frac{1}{q^2-m_2^2}-\frac{1}{q^2-m_1^2}\right) \cong i\rlap{/}{q}\left(\frac{m_2^2-m_1^2}{(q^2)^2}\right) = \frac{i\rlap{/}{q}}{(q^2)^2} 4m_M m_D \quad . \tag{11}$$

In the last step in this relation, we have used Eqns. (7) and (8). We see that, despite the combination of a LH current and a RH one, the amplitude for $0\nu\beta\beta$ still requires neutrino mass. In fact, it is quadratic in neutrino mass,[2] being proportional to $m_M m_D$,[3] whereas the amplitude that results when a LH current acts at both leptonic vertices, as in the SM, is only linear in mass.

The reason that a LH-RH current combination leads to an amplitude quadratic in neutrino mass is made clear by Fig. 4. Since both currents are lepton-number conserving, the $\bar{\nu}$ created together with an electron by the current at the vertex on the left must be converted to a $\nu$ so that it can be absorbed to make another electron by the current at the vertex on the right. Hence, one Majorana mass $m_M$ must be inserted along the neutrino line. But this mass will also flip the handedness of the exchanged particle, turning it into a LH neutrino, with the wrong handedness to be absorbed by a RH current. Thus, another mass insertion is needed to flip the handedness back again. Since this second mass insertion must not at the same time turn the exchanged particle back into a $\bar{\nu}$, it must be a Dirac mass $m_D$.[3]

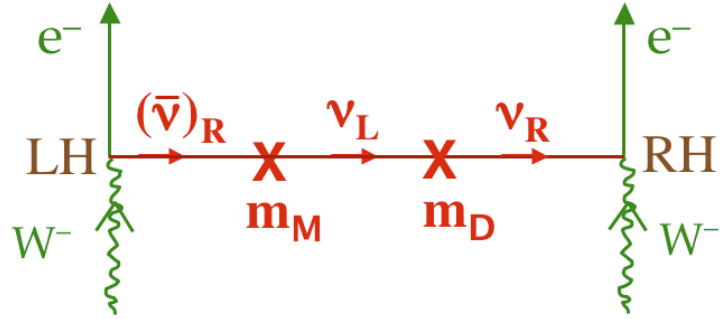

Figure 4. The particle-physics part of $0\nu\beta\beta$, with neutrino mass treated perturbatively, and a LH current acting at one vertex, but a RH one at the other, as indicated.

From this discussion, it is clear that if we go to higher order in neutrino mass and have more than two mass insertions in Fig. 4, there must be an odd number of insertions of $m_M$, to flip the lepton number, and also an odd number of insertions of $m_D$, so that there is no net flip of handedness. Thus, if the expression for $A_{LR}$ of Eqn. (11) is expanded in powers of $m_2$ and $m_1$, and the expansion is rewritten in terms of $m_M$ and $m_D$, every term must be of the form $m_M^{Odd\ Power} m_D^{Odd\ Power}$. It is trivial to see that this is indeed the case.

This simple toy model nicely illustrates the fact that, absent any lepton-number-nonconserving interactions, $0\nu\beta\beta$ requires Majorana neutrino mass. This mass is needed to introduce the lepton-number violation without which $0\nu\beta\beta$ cannot occur.

At this Symposium, we paid special tribute to the work of Frank Avignone, Ettore Fiorini, and Peter Rosen. In collaboration with Henry Primakoff, Peter Rosen was a pioneer of the whole field of double beta decay. A half-century ago, Primakoff and Rosen contemplated $0\nu\beta\beta$ arising from non-SM lepton-number-nonconserving interactions. Then Majorana neutrino mass is not needed. However, we have not yet seen any evidence for lepton-number-nonconserving interactions, but we do expect Majorana neutrino masses.

Frank Avignone and Ettore Fiorini have helped lead the way to the present very exciting point in the experimental quest for *0νββ*. Of course, we cannot know in advance what the rate for *0νββ* will prove to be, or even if the process occurs at all. But our present understanding of neutrino physics gives us good reasons to think that *0νββ* most likely does occur. It also allows us to make the rather plausible estimate that the $m_{\beta\beta}$ of Eqn. (4) lies in the range $2 \text{ meV} < m_{\beta\beta} < 40 \text{ meV}$,[4] which leads to lifetimes $\tau$ for *0νββ* in the rough range $10^{27} \text{ yr} < \tau < 10^{29} \text{ yr}$. Hopefully, the next one or two generations of *0νββ* experiments will achieve the required sensitivity, and we will see neutrinoless double beta decay at last.